\documentclass[conference]{IEEEtran}
\IEEEoverridecommandlockouts

\usepackage[letterpaper, left=1in, right=1in, bottom=1in, top=0.75in]{geometry}
\usepackage{graphicx}
\usepackage{url}
\usepackage{amsmath}
\usepackage{amsthm, amssymb}

\usepackage{tikz}
\usetikzlibrary{patterns}
\usetikzlibrary{shapes,arrows}
\usetikzlibrary{circuits.ee.IEC}
\usepackage[utf8]{inputenc}
\usepackage{cite}
\IEEEoverridecommandlockouts \IEEEpubid{\makebox[\columnwidth]{ 978-1-5386-3531-5/17/\$31.00~\copyright~2017 IEEE \hfill} \hspace{\columnsep}\makebox[\columnwidth]{ }}
\renewcommand{\baselinestretch}{0.95}
\begin{document}

\title{Uplink and Downlink Transceiver Design for OFDM with Index Modulation in Multi-user Networks \thanks{The work of  Merve Yüzgeçcioğlu has received partly funding from the European Union's Horizon 2020 research and innovation programme under the Marie Sklodowzka-Curie grant agreement No 641985}}

\author{\IEEEauthorblockN{Merve Yüzgeçcioğlu and Eduard Jorswieck}
\IEEEauthorblockA{Communications Theory, Communications Laboratory\\
Dresden University of Technology, 
D-01062 Dresden, Germany\\
Email: \{merve.yuzgeccioglu,\,eduard.jorswieck\}@tu-dresden.de}}

\maketitle

\begin{abstract}
A new modulation scheme called OFDM with index modulation (OFDM-IM) is introduced recently. This scheme allows to transmit additional bits by mapping a part of incoming bit stream to the indices of the subcarriers. In this work, performance of OFDM-IM in multi-user networks for uplink and downlink scenario is studied. For both scenarios, novel base station designs are introduced in order to overcome the inter-user-interference (IUI). Simulation results show that OFDM-IM outperforms the classical OFDM in multi-user networks and IUI is eliminated successfully even in large networks. 
\end{abstract}

\IEEEpeerreviewmaketitle

\section{Introduction}

One of the key requirements of future networks is the high spectral efficiency.  To be able to satisfy this requirement IM application is introduced for OFDM in  \cite{Abu-alhiga2009} which allows to transmit additional bits with subcarrier indices. In this scheme, the number of active OFDM subcarriers varies according to incoming bit stream and the data is transmitted only on the selected subcarriers. An enhanced subcarrier index modulation OFDM (ESIM-OFDM) scheme has been proposed in \cite{Tsonev2011} which can operate without requiring excess subcarriers. However, this scheme requires higher order modulations to reach the same spectral efficiency as that of the classical OFDM. 

A more flexible structure which is called OFDM-IM is introduced in \cite{Basar2012,Basar2013}. In this scheme, the number of active subcarriers are predefined and the indices of subcarriers are chosen according to incoming bits.  The extension of OFDM-IM to MIMO systems is studied in \cite{Basar2015} and different detection schemes are introduced.  In \cite{Gong2013}, IM is developed as a new way of vector modulation for interleaved frequency division multiple access (IFDMA) systems. It is shown that the combination of subcarrier index modulation (SIM) and IFDMA can have the advantage of both low peak-to-average power ratio (PAPR) and better system performance. A new subcarrier grouping method for OFDM-IM is proposed in \cite{Wen2016} and shown that OFDM-IM with interleaved grouping can achieve SNR gain over classical OFDM under a low-order alphabet input. A tight upper bound on bit error rate (BER) of OFDM-IM is given in \cite{Ko2014}. Complementary cumulative distribution function (CCDF) for PAPR is studied and shown that OFDM-IM has significantly better performance than classical OFDM. Further improvement is achieved by applying interleaving method to OFDM-IM \cite{Xiao2014}. A generalized space-frequency index modulation (GSFIM), which is a promising modulation scheme  that uses both spatial domain and frequency domain to encode bits through indexing is introduced in \cite{Datta2016,Chakrapani2015} and low complexity encoding and detection schemes are proposed. Another generalization for OFDM-IM are proposed in \cite{Fan2014,Fan2015}.  

Additional to these extensive works, in \cite{Zhu2016,Zhu2016-2} performance of SIM-OFDM is studied for multi-user networks. In these works, the number of active subcarriers are related to the modulation order which limits the number of bits that can be transmitted at a channel use and the resulting system design strongly depends on the modulation order. In \cite{Zhu2016}, an iterative algorithm is introduced to detect incoming bits at the receiver side and in \cite{Zhu2016-2}, previous work is extended for massive MIMO systems with imperfect channel state information (CSI).

In this work, uplink and downlink error performance of OFDM-IM scheme in multi-user networks are studied. The number of active subcarriers are predefined at the system and OFDM-IM block is generated accordingly. Since the number of active subcarriers are independent from the modulation order, OFDM-IM provides a flexible design in contrast to SIM-OFDM. While it is necessary to increase the modulation order to achieve higher spectrum efficiency (SE) in SIM-OFDM, better SE can be achieved with lower modulation order with different number of active subcarriers with OFDM-IM. Furthermore, simulation results show that the OFDM-IM scheme has better performance than both the classical OFDM and SIM-OFDM.

The rest of the paper is organized as follows. In Sec.~\ref{sec:SystemModel}, system models of OFDM-IM in multi-user network are introduced for both uplink and downlink scenario. In the same section, the SE and PAPR are calculated and compared to the state-of-the-art schemes. In Sec.~\ref{sec:NumericalResults}, error performance of the system is investigated and finally in Sec.~\ref{sec:Conclusion} the paper is concluded.

\section{System Model}\label{sec:SystemModel}

In classical OFDM scheme, there are $N_{tot}$ available subcarriers and all the subcarriers are used in order to transmit $N_{tot}\log_2 M$ bits at a channel use, where $M$ is the modulation order. Unlike the conventional one, in OFDM-IM scheme, $K_{tot}$ out of $N_{tot}$ available subcarriers are chosen to be active according to the incoming bit stream at each channel use.  $M$-ary modulated symbols are transmitted on these $K_{tot}$ subcarriers while $N_{tot}-K_{tot}$ subcarriers remain idle and the location of the active subcarriers convey additional information. These subcarriers are divided into $G$ groups in order to have a feasible receiver structure which is explained in Sec.~\ref{sec:Uplink} and Sec.~\ref{sec:Downlink} for uplink and downlink transmission, respectively. Resulting number of available and active subcarriers in each group is $N = N_{tot}/G$ and $K = K_{tot}/G$. With this scheme,  $b = b_1 + b_2$ bits are transmitted at a channel use where  $b_1 = G\lfloor \log_2 {{N}\choose{K}}\rfloor$ by IM part and  $b_2 = GK\log_2M$ as $M$-ary modulated symbols. 

In the following subsections, the system model of OFDM-IM for both uplink and downlink transmission in multi-user networks are introduced. The base station (BS) design to deal with the (IUI) for these systems are explained in detail.

\subsection{Uplink Transmission}\label{sec:Uplink}

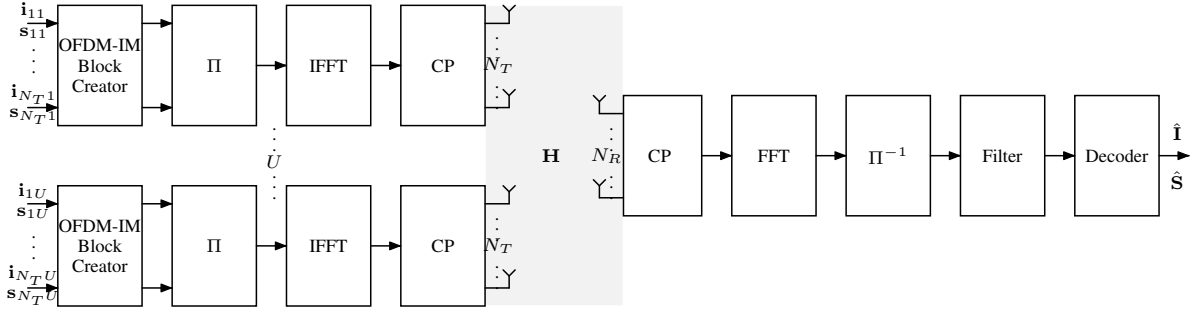
\begin{figure*}
\centering
\usetikzlibrary{circuits.ee.IEC}
\usetikzlibrary{arrows,decorations.markings}


\begin{tikzpicture}[circuit ee IEC, scale=0.8, line width=0.02cm, line cap=round, font=\scriptsize]

	\filldraw[fill=black!5!white, draw=black!0!white] (-4,2.5)rectangle (-1.7,-2.5);
	\node[align=center] at (-2.9,0) {$\mathbf{H}$};


	
	\node[align=center] at (-11.5,2.25) {$\mathbf{i}_{11}$\\$\mathbf{s}_{11}$};
	
	\node[align=center] at (-11.5,1.7) {$\vdots$};
	
	\node[align=center] at (-11.5,0.85) {$\mathbf{i}_{N_T1}$\\$\mathbf{s}_{N_T1}$};
	
	\draw (-11.6,2.2) -- (-11.1,2.2);
	\node[current direction={rotate = 0}] at (-11.2,2.2) {};
	
	\draw (-11.6,0.8) -- (-11.1,0.8);
	\node[current direction={rotate = 0}] at (-11.2,0.8) {};
	
	\draw (-11.1,2.5) rectangle (-9.7,0.5);
	\node[align=center] at (-10.4,1.5) {OFDM-IM\\Block\\Creator};
	
	\draw (-9.7,2.2) -- (-9.2,2.2);
	\node[current direction={rotate = 0}] at (-9.3,2.2) {};
	
	\draw (-9.7,0.8) -- (-9.2,0.8);
	\node[current direction={rotate = 0}] at (-9.3,0.8) {};
	
	\draw (-9.2,2.5) rectangle (-7.8,0.5);
	\node[align=center] at (-8.5,1.5) {$\Pi$};

	\draw (-7.8,1.5) -- (-7.3,1.5);
	\node[current direction={rotate = 0}] at (-7.4,1.5) {};
	
	\draw (-7.3,2.5) rectangle (-5.9,0.5);
	\node[align=center] at (-6.6,1.5) {IFFT};
	
	\draw (-5.9,1.5) -- (-5.4,1.5);
	\node[current direction={rotate = 0}] at (-5.5,1.5) {};
	
	\draw (-5.4,2.5) rectangle (-4,0.5);
	\node[align=center] at (-4.7,1.5) {CP};
	
	\draw (-4,2.2) -- (-3.6,2.2);
	\draw (-3.6,2.2) -- (-3.6,2.4);
	\draw (-3.6,2.4) -- (-3.7,2.5);
	\draw (-3.6,2.4) -- (-3.5,2.5);
	
	\node[align=center] at (-3.8,1.9) {$\vdots$};
	\node[align=center] at (-3.78,1.5) {$N_T$};
	\node[align=center] at (-3.8,1.1) {$\vdots$};
	
	\draw (-4,0.8) -- (-3.6,0.8);
	\draw (-3.6,0.8) -- (-3.6,1);
	\draw (-3.6,1) -- (-3.7,1.1);
	\draw (-3.6,1) -- (-3.5,1.1);
	
	\node[align=center] at (-7.5,0.4) {$\vdots$};
	\node[align=center] at (-7.5,-0.1) {$U$};
	\node[align=center] at (-7.5,-0.4) {$\vdots$};
	
	
	\node[align=center] at (-11.5,-0.75) {$\mathbf{i}_{1U}$\\$\mathbf{s}_{1U}$};
	
	\node[align=center] at (-11.5,-1.4) {$\vdots$};
	
	\node[align=center] at (-11.5,-2.15) {$\mathbf{i}_{N_TU}$\\$\mathbf{s}_{N_TU}$};
	
	\draw (-11.6,-0.8) -- (-11.1,-0.8);
	\node[current direction={rotate = 0}] at (-11.2,-0.8) {};
	
	\draw (-11.6,-2.2) -- (-11.1,-2.2);
	\node[current direction={rotate = 0}] at (-11.2,-2.2) {};
	
	\draw (-11.1,-0.5) rectangle (-9.7,-2.5);
	\node[align=center] at (-10.4,-1.5) {OFDM-IM\\Block\\Creator};
	
	\draw (-9.7,-0.8) -- (-9.2,-0.8);
	\node[current direction={rotate = 0}] at (-9.3,-0.8) {};
	
	\draw (-9.7,-2.2) -- (-9.2,-2.2);
	\node[current direction={rotate = 0}] at (-9.3,-2.2) {};
	
	\draw (-9.2,-0.5) rectangle (-7.8,-2.5);
	\node[align=center] at (-8.5,-1.5) {$\Pi$};

	\draw (-7.8,-1.5) -- (-7.3,-1.5);
	\node[current direction={rotate = 0}] at (-7.4,-1.5) {};
	
	\draw (-7.3,-0.5) rectangle (-5.9,-2.5);
	\node[align=center] at (-6.6,-1.5) {IFFT};
	
	\draw (-5.9,-1.5) -- (-5.4,-1.5);
	\node[current direction={rotate = 0}] at (-5.5,-1.5) {};
	
	\draw (-5.4,-0.5) rectangle (-4,-2.5);
	\node[align=center] at (-4.7,-1.5) {CP};
	
	\draw (-4,-0.8) -- (-3.6,-0.8);
	\draw (-3.6,-0.8) -- (-3.6,-0.6);
	\draw (-3.6,-0.6) -- (-3.7,-0.5);
	\draw (-3.6,-0.6) -- (-3.5,-0.5);
	
	\node[align=center] at (-3.8,-1.9) {$\vdots$};
	\node[align=center] at (-3.78,-1.5) {$N_T$};
	\node[align=center] at (-3.8,-1.1) {$\vdots$};
	
	\draw (-4,-2.2) -- (-3.6,-2.2);
	\draw (-3.6,-2.2) -- (-3.6,-2);
	\draw (-3.6,-2) -- (-3.7,-1.9);
	\draw (-3.6,-2) -- (-3.5,-1.9);


	\draw (-2.1,0.7) -- (-1.7,0.7);
	\draw (-2.1,0.7) -- (-2.1,0.9);
	\draw (-2.1,0.9) -- (-2,1);
	\draw (-2.1,0.9) -- (-2.2,1);
	
	\node[align=center] at (-1.9,0.4) {$\vdots$};
	\node[align=center] at (-1.98,0) {$N_R$};
	\node[align=center] at (-1.9,-0.4) {$\vdots$};
	
	\draw (-2.1,-0.7) -- (-1.7,-0.7);
	\draw (-2.1,-0.7) -- (-2.1,-0.5);
	\draw (-2.1,-0.5) -- (-2,-0.4);
	\draw (-2.1,-0.5) -- (-2.2,-0.4);

	\draw (-1.7,1) rectangle (-0.4,-1);
	\node[align=center] at (-1.1,0) {CP};
	
	\draw (-0.4,0) -- (0.1,0);
	\node[current direction={rotate = 0}] at (0,0) {};

	\draw (0.1,1) rectangle (1.5,-1);
	\node[align=center] at (0.8,0) {FFT};
	
	\draw (1.5,0) -- (2,0);
	\node[current direction={rotate = 0}] at (1.9,0) {};
	
	\draw (2,1) rectangle (3.4,-1);
	\node[align=center] at (2.7,0) {$\Pi^{-1}$};
	
	\draw (3.4,0) -- (3.9,0);
	\node[current direction={rotate = 0}] at (3.8,0) {};
	
	\draw (3.9,1) rectangle (5.3,-1);
	\node[align=center] at (4.6,0) {Filter};
	
	\draw (5.3,0) -- (5.8,0);
	\node[current direction={rotate = 0}] at (5.7,0) {};
	
	\draw (5.8,1) rectangle (7.2,-1);
	\node[align=center] at (6.5,0) {Decoder};
	
	\draw (7.2,0) -- (7.7,0);
	\node[current direction={rotate = 0}] at (7.6,0) {};
	\node[align=center] at (7.5,0.4) {$\hat{\mathbf{I}}$};
	\node[align=center] at (7.5,-0.4) {$\hat{\mathbf{S}}$};
	
\end{tikzpicture}
\caption{Block diagram of OFDM-IM uplink scenario in multi-user networks}
\label{fig:UplinkBD}
\end{figure*}

The transceiver block diagram for uplink transmission is shown in Fig.~\ref{fig:UplinkBD}. At this structure, there are $U$ users with $N_T$ transmit antennas who communicate with the BS with $N_R$ receive antennas where $N_R \geq UN_T$.  It is assumed that BS has CSI and users are not aware of the channel statistics. At each user, $b$ bits enter to each transmit antenna chain and these bits are divided into $G = N_{tot}/N$ groups such that $b = Gp$. Furthermore, $p$ bits are divided into two parts according to the modulation order $M$ and the number of active subcarriers $K$. According to $p_1$ bits,  the combination of  $K$ subcarriers is selected from the look-up table, $\mathbf{i}_{tu}^g = [i_{tu}^g(1) \quad i_{tu}^g(2) \quad \dots \quad i_{tu}^g(K)]^T$ where  $t = 1,\dots, N_T$, $u = 1,\dots, U$ and $g = 1,\dots, G$. Here, $i_{tu}^g(k) = 1,\dots,N$ is the selected subcarrier index of $g$-th group at $u$-th user to transmit from $t$-th transmit antenna. Note that, when $p_1$ bits are not the order of $2$, look-up table will be truncated and the decoding of the active subcarrier indices must be adapted accordingly. Furthermore, the remaining $p_2$ bits are used to generate $M$-ary modulated data, $\mathbf{s}_{tu}^g = [s_{tu}^g(1) \quad s_{tu}^g(2) \quad \dots \quad s_{tu}^g(K)]^T$ to transmit on selected subcarriers. The resulting subcarrier indices and the modulated symbols for all groups are $\mathbf{i}_{tu} = [(\mathbf{i}_{tu}^g)^T \quad (\mathbf{i}_{tu}^g)^T \quad \dots \quad (\mathbf{i}_{tu}^g)^T]^T$ and $\mathbf{s}_{tu} = [(\mathbf{s}_{tu}^g)^T \quad (\mathbf{s}_{tu}^g)^T \quad \dots \quad (\mathbf{s}_{tu}^g)^T]^T$, respectively, where $\mathbf{i}_{tu}$ and $\mathbf{s}_{tu}$ are $\in \mathbb{C}^{K_{tot}\times 1}$.

After the selection procedure is completed, modulated data are assigned to the active subcarriers. Firstly, the frequency domain OFDM-IM symbol of $g-$th group $\mathbf{x}_{tu}^g = [{x}_{tu}^g(1) \quad {x}_{tu}^g(2) \quad \dots \quad {x}_{tu}^g(N)]^T$ is generated. Then, in order to be sure that the symbols at each subcarrier are transmitted through uncorrelated channels, interleaved grouping is employed to generate final OFDM-IM block $\mathbf{x}_{tu}$, where $\mathbf{x}_{tu}[1, \dots,  g, \dots,  g+G, \dots,  g+(N-1)G,  \dots, N_{tot}]^T = \mathbf{x}_{tu}^g[1, 2, \dots,  N]^T$ for $t = 1,\dots, N_T$, $u = 1,\dots, U$ and $g = 1,\dots, G$. Note that, $G(N-K)$ elements that are the indices of the subcarriers were not selected as active are zero in $\mathbf{x}_{tu}$.  After reorganizing the frequency domain symbol $\mathbf{x}_{tu}$, $N_{tot}$-point IFFT operation is employed. By adding $N_{CP}$ length cyclic prefix, $N_{tot}+N_{CP}$ length OFDM-IM block is transmitted from each transmit antenna over an $L$-tap frequency-selective Rayleigh fading channel  from each user.  Input-output relationship of the system at frequency domain for each subcarrier is as follows

\begin{equation}
\mathbf{y}_n =  \sum_{u=1}^U\mathbf{H}_{nu}\mathbf{x}_{nu} + \mathbf{w}_n, \quad n = 1\dots N,
\end{equation}

where $\mathbf{x}_{nu} \in \mathbb{C}^{N_T \times 1}$ is the transmitted OFDM-IM symbol from $u$-th user, $\mathbf{H}_{nu}\in \mathbb{C}^{N_R \times N_T}$ is the effective channel matrix for $u$-th user with $\mathcal{CN}(0,1)$ distribution and  $\mathbf{w}_n \in \mathbb{C}^{N_R \times 1}$ is the AWGN with $\mathcal{CN}(0,\sigma^2)$ distribution, respectively. 

At the base station, cyclic prefix is removed and $N_{tot}$-point FFT is employed. Following regrouping of the frequency domain signal, minimum mean square error (MMSE) filtering is applied to eliminate the IUI and successfully reconstruct  the transmitted symbol. The filtered signal at each group is as follows

\begin{equation}
\tilde{\mathbf{y}}_n^g = \mathbf{W}_n^g \mathbf{y}_n^g = \mathbf{W}_n^g \mathbf{H}_n^g\mathbf{x}_n^g + \mathbf{W}_n^g \mathbf{w}_n^g,
\end{equation}

where $\mathbf{W}_n^g = ((\mathbf{H}_n^g)^H\mathbf{H}_n^g + \mathbf{I}_{UN_T}/\rho)^{-1}(\mathbf{H}_n^g)^H$ is the MMSE filter of $g$-th group and $n$-th subcarrier such that  $\mathbf{H}_n^g = [\mathbf{H}_{n1}^g\quad \mathbf{H}_{n2}^g\quad \dots \quad \mathbf{H}_{nU}^g] \in \mathbb{C}^{N_R \times UN_T}$, $\rho = Up_s/\sigma^2$ is the SNR of the system and $p_s$ is the transmit power of an OFDM-IM block. The input signal $\mathbf{x}_n^g = [(\mathbf{x}_{n1}^g)^T\quad (\mathbf{x}_{n2}^g)^T\quad \dots \quad (\mathbf{x}_{nU}^g)^T]^T$ is $UN_T\times 1$ vector that contains the data from all the users. 

After applying MMSE filtering to each subcarrier, the resulting  $\tilde{\mathbf{y}}_n^g\in \mathbb{C}^{UN_T\times 1}$ symbols are rearranged such that $\tilde{\mathbf{X}}^g  = [\tilde{\mathbf{X}}_1^g \quad \tilde{\mathbf{X}}_2^g \quad  \dots \quad \tilde{\mathbf{X}}_U^g]$ where $\tilde{\mathbf{X}}_u^g$ is $\mathbb{C}^{N\times N_T}$ for $u = 1,\dots,U$. Once we have the filtered matrix $\tilde{\mathbf{X}}^g$, it is easy to detect the active indices. Assume  $\mathbf{\Phi} \in \mathbb{Z}^{2^{p_1}\times K}$ is the matrix that contains all the possible index combinations of truncated look-up table. An example of $\mathbf{\Phi}$ is given for $N = 4$ and $K = 2$

\begin{equation}
\mathbf{\Phi} = 
 \begin{bmatrix}
    1\quad 1\quad 1\quad 2\\
    2\quad 3\quad 4\quad 3
\end{bmatrix}^T.
\end{equation}

By using such a look-up table, the decision metric is calculated for each combination 

\begin{equation}
d(l) = \sum_{k=1}^K |\tilde{\mathbf{x}}_{tu}^g(\mathbf{\Phi}(l,k))|,
\end{equation}

where $l = 1,\dots,2^{p_1}$, $t = 1,\dots,N_T$, $u = 1,\dots,U$ and $g = 1,\dots,G$. Once we have the decision metric $\mathbf{d}$, the maximum entry of this metric that also indicates the active subcarrier combination is found as $\hat{l} = \text{arg}\displaystyle\max_{l}{\mathbf{d}(l)}$. Finally, $\hat{l}$ is mapped to $\mathbf{\Phi}$ and the active subcarriers are detected as follows

\begin{equation}
\hat{\mathbf{i}}_{tu}^g = \mathbf{\Phi}(\hat{l},:).
\end{equation}

After detection of the active subcarrier indices, the symbols on these subcarriers are collected as $\hat{\mathbf{x}}_{tu}^g = \tilde{\mathbf{x}}_{tu}^g(\hat{\mathbf{i}}_{tu}^g)$. Here, $\hat{\mathbf{i}}_{tu}^g$ and $\hat{\mathbf{x}}_{tu}^g$ are $\mathbb{C}^{K\times 1}$ vectors that contain indices of $K$ active subcarriers and $M$-ary modulated data transmitted from $t$-th transmit antenna of $u$-th user, respectively.  From this point on, $\hat{\mathbf{s}}_{tu}^g$ can be found by $M$-ary demodulation of  $\hat{\mathbf{x}}_{tu}^g$ vector. The resulting detected indices and $M$-ary modulated data of all groups from all the users are $[\hat{\mathbf{I}}_1,\dots,\hat{\mathbf{I}}_U]$ and $[\hat{\mathbf{S}}_1,\dots,\hat{\mathbf{S}}_U]$, respectively, where $\hat{\mathbf{I}}_u$ and $\hat{\mathbf{S}}_u$ are $\in \mathbb{C}^{K_{tot}\times N_T}$ for $u = 1,\dots,U$.

\subsection{Downlink Transmission}\label{sec:Downlink}

\begin{figure*}
\centering
\usetikzlibrary{circuits.ee.IEC}
\usetikzlibrary{arrows,decorations.markings}


\begin{tikzpicture}[circuit ee IEC, scale=0.8, line width=0.02cm, line cap=round, font=\scriptsize]

	\filldraw[fill=black!5!white, draw=black!0!white] (-2.1,1)rectangle (0.4,-1);
	\node[align=center] at (-0.85,0) {$\mathbf{H}_u$};


	
	\node[align=center] at (-11.5,0.7) {$\mathbf{I}_1$\\$\mathbf{S}_1$};
	\node[align=center] at (-11.5,-0.7) {$\mathbf{I}_U$\\$\mathbf{S}_U$};
	
	\node[align=center] at (-11.5,0.2) {$\vdots$};
	
	\draw (-11.6,0.7) -- (-11.1,0.7);
	\node[current direction={rotate = 0}] at (-11.2,0.7) {};
	
	\draw (-11.6,-0.7) -- (-11.1,-0.7);
	\node[current direction={rotate = 0}] at (-11.2,-0.7) {};
	
	\draw (-11.1,1) rectangle (-9.7,-1);
	\node[align=center] at (-10.4,0) {OFDM-IM\\Block\\Creator};
	
	\draw (-9.7,0.7) -- (-9.2,0.7);
	\node[current direction={rotate = 0}] at (-9.3,0.7) {};
	
	\draw (-9.7,-0.7) -- (-9.2,-0.7);
	\node[current direction={rotate = 0}] at (-9.3,-0.7) {};
	
	\draw (-9.2,1) rectangle (-7.8,-1);
	\node[align=center] at (-8.5,0) {$\Pi$};

	\draw (-7.8,0) -- (-7.3,0);
	\node[current direction={rotate = 0}] at (-7.4,0) {};
	
	\draw (-7.3,1) rectangle (-5.9,-1);
	\node[align=center] at (-6.6,0) {Precoder};
	
	\draw (-5.9,0) -- (-5.4,0);
	\node[current direction={rotate = 0}] at (-5.5,0) {};
	
	\draw (-5.4,1) rectangle (-4,-1);
	\node[align=center] at (-4.7,0) {IFFT};
	
	\draw (-4,0) -- (-3.5,0);
	\node[current direction={rotate = 0}] at (-3.6,0) {};
	
	\draw (-3.5,1) rectangle (-2.1,-1);
	\node[align=center] at (-2.8,0) {CP};
	
	\draw (-2.1,0.7) -- (-1.7,0.7);
	\draw (-1.7,0.7) -- (-1.7,0.9);
	\draw (-1.7,0.9) -- (-1.8,1);
	\draw (-1.7,0.9) -- (-1.6,1);
	
	\node[align=center] at (-1.9,0.4) {$\vdots$};
	\node[align=center] at (-1.88,0) {$N_T$};
	\node[align=center] at (-1.9,-0.4) {$\vdots$};
	
	\draw (-2.1,-0.7) -- (-1.7,-0.7);
	\draw (-1.7,-0.7) -- (-1.7,-0.5);
	\draw (-1.7,-0.5) -- (-1.8,-0.4);
	\draw (-1.7,-0.5) -- (-1.6,-0.4);


	\draw (0,0.7) -- (0.4,0.7);
	\draw (0,0.7) -- (0,0.9);
	\draw (0,0.9) -- (0.1,1);
	\draw (0,0.9) -- (-0.1,1);
	
	\node[align=center] at (0.2,0.4) {$\vdots$};
	\node[align=center] at (0.17,0) {$N_R$};
	\node[align=center] at (0.2,-0.4) {$\vdots$};
	
	\draw (0,-0.7) -- (0.4,-0.7);
	\draw (0,-0.7) -- (0,-0.5);
	\draw (0,-0.5) -- (0.1,-0.4);
	\draw (0,-0.5) -- (-0.1,-0.4);

	\draw (0.4,1) rectangle (1.8,-1);
	\node[align=center] at (1.1,0) {CP};
	
	\draw (1.8,0) -- (2.3,0);
	\node[current direction={rotate = 0}] at (2.2,0) {};

	\draw (2.3,1) rectangle (3.7,-1);
	\node[align=center] at (3,0) {FFT};
	
	\draw (3.7,0) -- (4.2,0);
	\node[current direction={rotate = 0}] at (4.1,0) {};
	
	\draw (4.2,1) rectangle (5.6,-1);
	\node[align=center] at (4.9,0) {$\Pi^{-1}$};
	
	\draw (5.6,0) -- (6.1,0);
	\node[current direction={rotate = 0}] at (6.0,0) {};
	
	\draw (6.1,1) rectangle (7.5,-1);
	\node[align=center] at (6.8,0) {Decoder};
	
	\draw (7.5,0) -- (8.0,0);
	\node[current direction={rotate = 0}] at (7.9,0) {};
	
	\node[align=center] at (7.8,0) {$\hat{\mathbf{i}}_u$\\\\$\hat{\mathbf{s}}_u$};
	
\end{tikzpicture}
\caption{Block diagram of OFDM-IM downlink scenario in multi-user networks}
\label{fig:DownlinkBD}
\end{figure*}
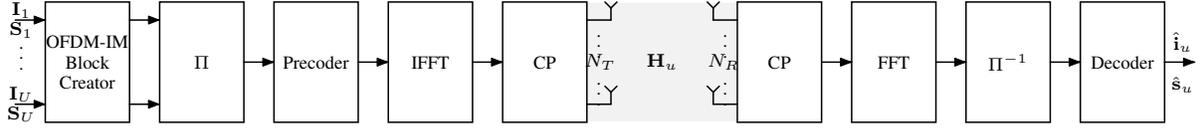

The transceiver block diagram for the downlink transmission is shown in Fig.~\ref{fig:DownlinkBD}. In this system, BS transmits information to $U$ users. BS has $N_T$ transmit antennas such that $N_T \geq UN_R$ and users have $N_R$ receive antennas. The procedure to generate OFDM-IM symbols $\mathbf{x}_{ru}^g$ for each group is same as the uplink model that is explained in detail in Sec.~\ref{sec:Uplink}, where $r = 1,\dots,N_R$, $u = 1,\dots,U$ and $g = 1,\dots,G$. OFDM-IM symbols are merged into $\mathbf{x}_{ru}\in \mathbb{C}^{N_{tot}\times 1}$ by interleaved grouping to build the OFDM-IM block up. 

The difference from the uplink transmission, at downlink case CSIT is available at the BS and users have only partial information on channel statistics. By taking advantage of the CSIT, MMSE-based precoder is employed in order to eliminate the IUI. The precoded frequency domain OFDM-IM symbol is 

\begin{equation}
\bar{\mathbf{x}}_n = \gamma_n \sum_{u=1}^U\mathbf{P}_{nu}\mathbf{x}_{nu}, \quad n = 1\dots N,
\end{equation}

where $\mathbf{x}_{nu} \in \mathbb{C}^{N_R\times 1}$ is the signal on $n$-th subcarrier to be transmitted to the $u$-th user. $\mathbf{P}_{nu} \in \mathbb{C}^{N_T\times N_R}$ is the precoding matrix designed to mitigate IUI such that 

\begin{equation}
\mathbf{P}_n =  \left (\mathbf{H}_n^H\mathbf{H}_n + \frac{\mathbf{I}_{UN_T}}{\rho}\right )^{-1}\mathbf{H}_n^H.
\end{equation}

Here, $\mathbf{P}_n = [\mathbf{P}_{n1} \quad \mathbf{P}_{n2} \quad \dots \quad \mathbf{P}_{nU}]$ is the precoding matrix for all users where $u = 1,\dots,U$ and $\gamma_n = \sqrt{\frac{U}{\text{trace}\{\mathbf{P}_n \mathbf{P}_n^H\}}}$ is the normalization coefficient. After this point, same procedure is applied  as in the uplink transmission.

The received signal at $u$-th user after removing the cyclic prefix and interleaved regrouping is

\begin{equation}
\tilde{\mathbf{y}}_{nu}^g  = \mathbf{H}_{nu}^g\bar{\mathbf{x}}_n^g + \mathbf{w}_{nu}^g,
\end{equation}

where $\bar{\mathbf{x}}_n^g \in \mathbb{C}^{N_T\times 1}$ is the transmitted symbol, $\mathbf{H}_{nu}^g \in \mathbb{C}^{N_R \times N_T}$ is the Rayleigh fading channel matrix between $u$-th user and the BS with  $\mathcal{CN}(0,1)$ distribution and  $\mathbf{w}_{nu}^g \in \mathbb{C}^{N_R \times 1}$ is the AWGN with $\mathcal{CN}(0,\sigma^2)$ distribution, respectively. 

Since the precoding is applied in order to eliminate IUI, the user only needs to calculate $\tilde{\mathbf{x}}_{nu}^g =  \tilde{\mathbf{y}}_{nu}^g/\gamma_n$ and then proceeds to decode the received OFDM-IM symbol. Note that, with this design, user only needs to know $\mathbf{\gamma}_n$ values, instead of whole channel matrix $\mathbf{H}_n \in \mathbb{C}^{N_T\times N_R}$ where $n = 1,\dots,N_{tot}$. Furthermore, the active subcarrier indices $\hat{\mathbf{i}}_{ru}^g$ and the $M$-ary symbols $\hat{\mathbf{s}}_{ru}^g$ on these subcarriers are detected as described in Sec.~\ref{sec:Uplink}, where $r = 1,\dots, N_R$, $u = 1,\dots, U$ and $g = 1,\dots, G$. 

\subsection{Performance Analysis}\label{sec:PerformanceAnalysis}

The upper bound of the SE for OFDM-IM scheme can be calculated as follows

\begin{equation}
SE_{OFDM-IM} = \frac{G(\lfloor \log_2 {{N}\choose{K}}\rfloor + K\log_2 M)}{N_{tot}+N_{CP}}.
\end{equation}

On the other hand, the upper bound of the SE for classical OFDM and SIM-OFDM are given below

\begin{equation}
SE_{OFDM} = \frac{N_{tot}\log_2 M}{N_{tot}+N_{CP}},
\end{equation}

\begin{equation}\label{eq:SE_SIM}
\begin{split}
SE_{SIM-OFDM} &=  \frac{G(\log_2 M + K \log_2 M)}{N_{tot}+N_{CP}} \\
															&=  \frac{N_{tot}\log_2 M}{N_{tot}+N_{CP}}.
\end{split}
\end{equation}

Since $K = M-1$ for the SIM-OFDM scheme where $M$ is the modulation order, it is seen from Eq.~\eqref{eq:SE_SIM} that the SE of the SIM-OFDM is equal to SE of the classical OFDM, $SE_{SIM-OFDM} = SE_{OFDM}$. On the contrary, the parameters $N$ and $K$ can be defined independently from the modulation order for the OFDM-IM scheme,  with certain parameters, it is possible to achieve better SE than both OFDM and SIM-OFDM schemes. As an example: a SIM-OFDM system with $N_{tot} = 128$, $N_{CP} = 8$, $M = 4$ the SE of SIM-OFDM is $SE_{SIM-OFDM} = 1.88$; on the other hand an OFDM-IM system with $N_{tot} = 128$, $N_{CP} = 8$, $N = 16$, $K = 12$, $M = 4$ the SE is  $SE_{OFDM-IM} = 2$.

Another performance metric in that the OFDM-IM is advantageous is PAPR. The PAPR values for all three schemes can be found as follows

\begin{equation}
PAPR_{OFDM-IM} = 10\log_{10} GK,
\end{equation}

\begin{equation}
PAPR_{SIM-OFDM} = 10\log_{10} GK,
\end{equation}

\begin{equation}
PAPR_{OFDM} = 10\log_{10} N_{tot}.
\end{equation}

It is clearly seen from above equations that $PAPR_{OFDM-IM} \leq PAPR_{SIM-OFDM} < PAPR_{OFDM}$.

\section{Numerical Results}\label{sec:NumericalResults}

\begin{figure}[!t]
\centering
\includegraphics[scale=0.4]{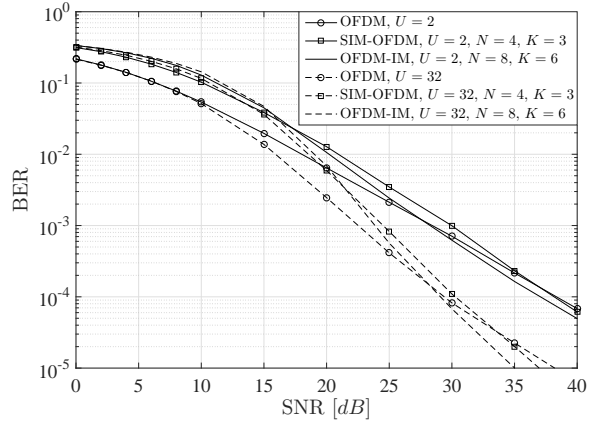}
\caption{BER comparison of uplink transmission for OFDM, SIM-OFDM and OFDM-IM in $2$- and $32$-single antenna user networks with $N_R = 2$ and $N_R = 32$, respectively.}
\label{fig:UL}
\end{figure}

In this section we have shown the performance of OFDM-IM for both uplink and downlink transmission. We considered the BER as performance metric. In all simulations, we assume that the total number of subcarriers is $N_{tot} = 128$, $N_{CP} = 16$ and $4$-QAM is considered to modulate the bits. Furthermore, $L = 8$ tap frequency selective Rayleigh fading channel is considered between each user and the BS.

In Fig.~\ref{fig:UL}, the uplink BER performance of the system is shown. We  compared the OFDM-IM scheme with classical OFDM and SIM-OFDM that is studied for multi-user scenario in \cite{Zhu2016}. In Fig.~\ref{fig:UL}, we consider the uplink scenario with $2$ and $32$ users. Each user is equipped with a single transmit antenna and the BS has $N_R = 2$ and $N_R = 32$ receive antennas, respectively. Since the number of active subcarriers are determined by the modulation order in the SIM-OFDM scheme, it is assumed that $N = 4$, $K = 3$. On the other hand, the number of total and active subcarriers in a group  are determined as $N = 8$ and $K = 6$ for the OFDM-IM scheme, respectively. It is seen in Fig.~\ref{fig:UL} that the OFDM-IM scheme achieves better performance than both classical OFDM and SIM-OFDM in both cases. Since only $GK$ subcarriers carry information on the spectrum, the average distance between $M$-ary symbols on the spectrum is larger than in the other two schemes. Furthermore, the interleaved grouping provides additional protection to modulated symbols against correlated channels. Note also that the selection of the available subcarriers and the active subcarriers in a group is independent from the modulation order. In contrary, in SIM-OFDM scheme, the number of subcarriers at a group is defined as $M$ that is also the modulation order and only one subcarrier is selected inactive at a channel use. This property of the scheme limits the system design while in OFDM-IM, $N$ and $K$ can be chosen freely according to the system requirements. This feature provides a flexible design and allows trade of between SE and PAPR.

\begin{figure}[!t]
\centering
\includegraphics[scale=0.4]{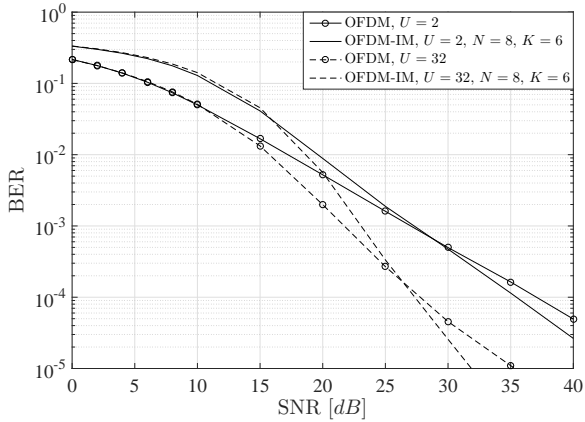}
\caption{BER comparison of downlink transmission for OFDM and OFDM-IM in $2$- and $32$-single antenna user networks with $N_T = 2$ and $N_T = 32$, respectively.}
\label{fig:DL}
\end{figure}

In Fig.~\ref{fig:DL}, we consider the downlink scenario with $2$ and $32$ users. Each user is equipped with a single transmit antenna and the BS has $N_T = 2$ and $N_T = 32$ transmit antennas, respectively. The number of total and active subcarriers in a group  are determined as $N = 8$ and $K = 6$ for OFDM-IM scheme, respectively. It is seen in Fig.~\ref{fig:DL} that the downlink performance of OFDM-IM scheme is similar to the uplink scenario. The MMSE-precoder successfully mitigates the IUI and OFDM-IM outperforms the classical OFDM for high SNR values. 

The poor performance of OFDM-IM in low SNR regime for both uplink and downlink cases can be explained with the higher erroneous detection of subcarrier indices. When the active subcarrier indices are not detected correctly at the receiver side, decoding of the modulated symbols on these detected subcarriers will also  be  erroneous. Another disadvantageous brought by the IM part of the scheme is the additional complexity at detection. There are intermediate  steps between filtering and $M$-ary demodulation in order to detect the active subcarrier indices. Despite the additional complexity, OFDM-IM have a better error performance than classical OFDM and SIM-OFDM schemes. Furthermore, OFDM-IM provides a flexible system model that allows us to design the system according to requirements such as high SE, low PAPR and low BER.

\section{Conclusion}\label{sec:Conclusion}

In this work, the OFDM-IM scheme has been studied for multi-user networks. System models for downlink and uplink scenarios have been introduced. Novel transceiver designs in order to eliminate IUI have been explained in detail for both of the scenarios. It has been shown that the OFDM-IM outperforms the classical OFDM in also multi-user networks. As future work, we believe that investigating the energy efficiency of the system is an interesting and timely research. 
\renewcommand{\baselinestretch}{0.9}
\bibliographystyle{IEEEtran}
\bibliography{article}

\begin{thebibliography}{10}
\providecommand{\url}[1]{#1}
\csname url@samestyle\endcsname
\providecommand{\newblock}{\relax}
\providecommand{\bibinfo}[2]{#2}
\providecommand{\BIBentrySTDinterwordspacing}{\spaceskip=0pt\relax}
\providecommand{\BIBentryALTinterwordstretchfactor}{4}
\providecommand{\BIBentryALTinterwordspacing}{\spaceskip=\fontdimen2\font plus
\BIBentryALTinterwordstretchfactor\fontdimen3\font minus
  \fontdimen4\font\relax}
\providecommand{\BIBforeignlanguage}[2]{{%
\expandafter\ifx\csname l@#1\endcsname\relax
\typeout{** WARNING: IEEEtran.bst: No hyphenation pattern has been}%
\typeout{** loaded for the language `#1'. Using the pattern for}%
\typeout{** the default language instead.}%
\else
\language=\csname l@#1\endcsname
\fi
#2}}
\providecommand{\BIBdecl}{\relax}
\BIBdecl

\bibitem{Abu-alhiga2009}
R.~Abu-alhiga and H.~Haas, ``Subcarrier-index modulation {OFDM},'' in
  \emph{IEEE 20th International Symposium on Personal, Indoor and Mobile Radio
  Communications}, Sep. 2009, pp. 177--181.

\bibitem{Tsonev2011}
D.~Tsonev, S.~Sinanovic, and H.~Haas, ``Enhanced subcarrier index modulation
  ({SIM}) {OFDM},'' in \emph{IEEE GLOBECOM Workshops (GC Wkshps)}, Dec. 2011,
  pp. 728--732.

\bibitem{Basar2012}
E.~Başar, Ümit Aygölü, E.~Panayırcı, and H.~V. Poor, ``Orthogonal
  frequency division multiplexing with index modulation,'' in \emph{IEEE Global
  Communications Conference (GLOBECOM)}, Dec. 2012, pp. 4741--4746.

\bibitem{Basar2013}
------, ``Orthogonal frequency division multiplexing with index modulation,''
  \emph{IEEE Transactions on Signal Processing}, vol.~61, no.~22, pp.
  5536--5549, Nov. 2013.

\bibitem{Basar2015}
E.~Başar, ``Multiple-input multiple-output {OFDM} with index modulation,''
  \emph{IEEE Signal Processing Letters}, vol.~22, no.~12, pp. 2259--2263, Dec.
  2015.

\bibitem{Gong2013}
L.~Gong, L.~Dan, S.~Feng, S.~Wang, Y.~Xiao, and S.~Li, ``Subcarrier-index based
  vector-modulated {IFDMA} systems,'' in \emph{International Conference on
  Communications, Circuits and Systems (ICCCAS)}, vol.~2, Nov. 2013, pp.
  19--21.

\bibitem{Wen2016}
M.~Wen, X.~Cheng, M.~Ma, B.~Jiao, and H.~V. Poor, ``On the achievable rate of
  {OFDM} with index modulation,'' \emph{IEEE Transactions on Signal
  Processing}, vol.~64, no.~8, pp. 1919--1932, Apr. 2016.

\bibitem{Ko2014}
Y.~Ko, ``A tight upper bound on bit error rate of joint {OFDM} and
  multi-carrier index keying,'' \emph{IEEE Communications Letters}, vol.~18,
  no.~10, pp. 1763--1766, Oct. 2014.

\bibitem{Xiao2014}
Y.~Xiao, S.~Wang, L.~Dan, X.~Lei, P.~Yang, and W.~Xiang, ``{OFDM} with
  interleaved subcarrier-index modulation,'' \emph{IEEE Communications
  Letters}, vol.~18, no.~8, pp. 1447--1450, Aug. 2014.

\bibitem{Datta2016}
T.~Datta, H.~S. Eshwaraiah, and A.~Chockalingam, ``Generalized
  space-and-frequency index modulation,'' \emph{IEEE Transactions on Vehicular
  Technology}, vol.~65, no.~7, pp. 4911--4924, Jul. 2016.

\bibitem{Chakrapani2015}
B.~Chakrapani, T.~L. Narasimhan, and A.~Chockalingam, ``Generalized
  space-frequency index modulation: Low-complexity encoding and detection,'' in
  \emph{IEEE Globecom Workshops}, Dec. 2015, pp. 1--6.

\bibitem{Fan2014}
R.~Fan, Y.~J. Yu, and Y.~L. Guan, ``Orthogonal frequency division multiplexing
  with generalized index modulation,'' in \emph{IEEE Global Communications
  Conference}, Dec. 2014, pp. 3880--3885.

\bibitem{Fan2015}
------, ``Generalization of orthogonal frequency division multiplexing with
  index modulation,'' \emph{IEEE Transactions on Wireless Communications},
  vol.~14, no.~10, pp. 5350--5359, Oct. 2015.

\bibitem{Zhu2016}
H.~Zhu, W.~Wang, Q.~Huang, and X.~Gao, ``Subcarrier index modulation {OFDM} for
  multiuser {MIMO} systems with iterative detection,'' in \emph{IEEE
  International Symposium on Personal, Indoor, and Mobile Radio
  Communications}, Sep. 2016, pp. 1--6.

\bibitem{Zhu2016-2}
------, ``Uplink transceiver for subcarrier index modulation {OFDM} in massive
  {MIMO} systems with imperfect channel state information,'' in
  \emph{International Conference on Wireless Communications Signal Processing},
  Oct. 2016, pp. 1--6.

\end{thebibliography}
\end{document}